\documentclass{article}
\newcommand{\bfr}{\begin{flushright}}
\newcommand{\efr}{\end{flushright}}
 
\begin{document}
\title{Multi-black hole solutions in cosmological
Einstein-Maxwell-dilaton theory
%\thanks{Presented at ...}%
}
\author{Takuya Maki\\
Department of Physics, Tokyo Metropolitan University,\\
Minami-ohsawa, Hachioji-shi, Tokyo 192-03, Japan\\
and\\
Kiyoshi Shiraishi\\
%\address{
Akita Junior College, Shimokitade-Sakura, Akita-shi, Akita 010, Japan
%}
}
\date{Class. Quantum Grav. {\bf 10} (1993) 2171--2178
}
\maketitle
\begin{abstract}
Adopting a simple ansatz, we find exact solutions to the
Einstein-Maxwell-dilaton equations, which stand for the multi-black
hole configuration with maximal charge in a cosmological metric and
dilaton field background driven by a cosmological term. 
\end{abstract}
%\PACS{}

In \cite{1}, one of the present authors has presented the
multi-centered solutions for the Einstein-Maxwell-dilaton system%
\footnote{%
Multi-black hole solutions with magnetic charges in string theory 
($a=1$) in four dimensions are shown in \cite{2}.
} in
arbitrary dimensions and analysed the low-energy scattering
of maximally charged dilatonic black holes by use of the metric in the
moduli space. In the present paper, we treat $(N+1)$-dimensional
cosmological solutions as a natural extension of our previous work,
describing the multi-black hole configuration in the same system with a
cosmological constant.

We believe that such solutions have some relevance to the very early
evolution of the Universe. In the very early era, all the dimensions
may be isotropic and the cosmological constant may not yet be
cancelled. Black holes may also be created by the initial density
fluctuations. The exact solutions which will be obtained here may
become a simple but useful tool for dealing with the early evolution of
the Universe.

Of course, the number of dimensions of the space is three and the
cosmological constant is known to be nearly zero in the present
universe. Moreover, the dilaton must be massive now. Inclusion of the
two mass scales, the compactification scale and the dilaton mass
scale, makes the simple, analytic solutions impossible in the
Einstein-Maxwell-dilaton system. Therefore we deal with the
uncompactified space and the massless dilaton. (We can obtain the usual
three-dimensional case when we set $N=3$ in our analysis.)

This work is partially influenced by the recent papers \cite{3,4}.

In order to study the role of dilaton, one can consider the action
with an arbitrary value for the dilaton couphg:
\begin{equation}
S=\int d^{N+1}x
\frac{\sqrt{-g}}{16\pi}\left[R-\frac{4}{N-1}(\nabla\phi)^2-
e^{-4a\phi/(N-1)}F^2-e^{4b\phi/(N-1)}\Lambda\right]
\label{eq1}
\end{equation}
where $R$ is the curvature scalar and $\Lambda$ is the cosmological
constant. The scalar $\phi$ is known as the dilaton, while $F_{\mu\nu}$
denotes the Maxwell field strength. The Newton constant in $N+1$
dimensions is normalized to unity here.

The constants $a$ and $b$ represent dilaton couplings to other
entities. The effective field theory of string theory corresponds to
the case with $a=b=1$, while the usual Einstein-Maxwell system is
obtained by setting $a$ and $b$ to be zero.

Varying the action (\ref{eq1}) gives the equations of motion:
\begin{eqnarray}
& &\nabla_\mu(e^{-4a\phi/(N-1)}F^{\mu\nu})=0
\label{eq2}\\
& &\nabla_\mu\nabla^\mu\phi=-\frac{a}{2}e^{-4a\phi/(N-1)}(F)^2+
\frac{b}{2}e^{4b\phi/(N-1)}\Lambda
\label{eq3}\\
&
&R_{\mu\nu}-\frac{1}{2}g_{\mu\nu}R=\frac{4}{N-1}\left[
\nabla_\mu\phi\nabla_\nu\phi-\frac{1}{2}g_{\mu\nu}(\nabla\phi)^2
\right]\nonumber
\\
&
&\qquad\qquad\quad\quad+e^{-4a\phi/(N-1)}\left[2F_{\mu\lambda}F_\nu{}^\lambda-
\frac{1}{2}g_{\mu\nu}(F)^2\right]-\frac{1}{2}e^{4b\phi/(N-1)}\Lambda\,.
\label{eq4}
\end{eqnarray}

First we consider a homogeneous, isotropic, charge-neutral solution to
the equations (\ref{eq2})-(\ref{eq4})%
\footnote{%
We consider expanding universe only throughout this paper, since
extension to contracting universe is straightforward.
}.
We impose a common ansatz on the
metric for a spatially-flat universe:
\begin{equation}
ds^2=-dt^2+R^2(t)\,d{\bf x}^2\,.
\label{eq5}
\end{equation}
We must remark that cosmological evolution of the scale factor depends
on the coupling of the dilaton to the cosmological constant. If 
$b=0$, i.e. there is no dilaton coupling to the cosmological constant,
the space becomes a de Sitter space, as is well known. As long as $b$
differs from zero, the space does not behave like the de Sitter space.

If such a space is vacant (i.e. there are no matter field and particle,
no electromagnetic field ($F_{\mu\nu}=0$)) and the dilaton field also
evolves homogeneously, the equations of motion become
\begin{eqnarray}
& &\ddot{\phi}+N\frac{\dot{R}}{R}{\dot{\phi}}
+\frac{b}{2}e^{4b\phi/(N-1)}\Lambda=0\\
\label{eq6}
& &N\frac{\ddot{R}}{R}+\frac{4}{N-1}\dot{\phi}^2-
\frac{1}{N-1}e^{4b\phi/(N-1)}\Lambda=0\\
\label{eq7}
& &\frac{\ddot{R}}{R}+(N-1)\left(\frac{\dot{R}}{R}\right)^2-
\frac{1}{N-1}e^{4b\phi/(N-1)}\Lambda=0
\label{eq8}
\end{eqnarray}
where the dot denotes the time derivative.

The solutions of these equations are known \cite{5}. The scale factor
$R(t)$ in the case of $b\ne 0$ is known to behave as:
\begin{equation}
R(t)=(t/t_0)^{1/b^2}
\label{eq9}
\end{equation}
where $t_0$ is a constant, and the dilaton field varies as
\begin{equation}
e^{4b\phi/(N-1)}=\Sigma\,t^{-2}\,.
\label{eq10}
\end{equation}
There is a relation among $b^2$, the constant $\Sigma$, and non-zero
$\Lambda$;
\begin{equation}
\frac{1}{b^2}\left(\frac{N}{b^2}-1\right)=\frac{\Sigma\Lambda}{N-1}\,.
\label{eq11}
\end{equation}
(In addition, a time-independent solution, $R(t)=constant$
and $\phi(t)=constant$ is
allowed if and only if $\Lambda$ vanishes.)

As for the universe which contains point sources with
charges and masses, the
following metric ansatz is considered as an extension of (\ref{eq5}):
\begin{equation}
ds^2=-U^{-2}({\bf x},t)\,dt^2+R^2(t)\,U^{2/(N-2)}({\bf x},t)\,d{\bf
x}^2.
\label{eq12}
\end{equation}
This form of metric is considered in \cite{3} for $N=3$.

The equations of motions can be reduced to the following set
of simultaneous differential equations by using the metric function
(\ref{eq12}):
\begin{eqnarray}
& &\frac{1}{R^NU^{2/(N-2)}}\left(R^NU^{2(N-1)/(N-2)}\dot{\phi}\right)
\dot{}+\frac{b}{2}e^{4b\phi/(N-1)}\Lambda\nonumber \\
& &\qquad\qquad\qquad-\frac{1}{R^NU^{2/(N-2)}}\partial_i\partial^i\phi+
\frac{ae^{4a\phi/(N-1)}(E_k)^2}{R^{2(N-1)}U^{2(N-1)/(N-2)}}=0
\label{eq13}\\
&
&\frac{N}{N-2}\frac{\ddot{U}}{U}+\frac{N}{(N-2)^2}\left(
\frac{\dot{U}}{U}\right)^2+N\frac{\ddot{R}}{R}
+\frac{N^2}{N-2}\frac{\dot{R}}{R}\frac{\dot{U}}{U}+\frac{4}{N-1}
\dot{\phi}^2\nonumber \\
& &\qquad\qquad-\frac{e^{4b\phi/(N-1)}\Lambda}{(N-1)U^2}
+\frac{1}{R^NU^{2(N-1)/(N-2)}}\partial_i
\left(\frac{\partial^iU}{U}\right)\nonumber \\
& &\qquad\qquad\qquad\qquad\qquad+\frac{2(N-2)}{N-1}
\frac{e^{4a\phi/(N-1)}(E_k)^2/U^2}{R^{2(N-1)}U^{2(N-1)/(N-2)}}=0
\label{eq14}\\
&
&[U^2(R^2U^{2/(N-2)})\dot{}]\dot{}+(N-2)\frac{\dot{R}}{R}U^2
(R^2U^{2/(N-2)})\dot{}\nonumber
\\ & &\qquad\qquad\qquad
-\frac{2R^2U^{2/(N-2)}e^{4b\phi/(N-1)}\Lambda}{N-1}
-\frac{2}{N-2}\partial_i
\left(\frac{\partial^iU}{U}\right)\nonumber \\ &
&\qquad\qquad\qquad\qquad\qquad\qquad\qquad
-\frac{4}{N-1}\frac{e^{4a\phi/(N-1)}(E_k)^2}{R^{2(N-2)}U^{2}}=0
\label{eq15}\\
& &\frac{N-1}{N-2}\frac{\partial_iU}{U}\frac{\partial_jU}{U}
+\frac{4}{N-1}{\partial_i\phi}{\partial_j\phi}-
\frac{2e^{4a\phi/(N-1)}}{R^{2(N-2)}U^2}E_iE_j=0
\label{eq16}\\
& &\frac{N-1}{N-2}\partial_i\left(\frac{\dot{U}}{U}\right)
+\frac{N-1}{N-2}\frac{\partial_iU}{U}\frac{\dot{U}}{U}
+(N-1)\frac{\partial_iU}{U}\frac{\dot{R}}{R}
+\frac{4}{N-1}{\partial_i\phi}\dot{\phi}=0
\label{eq17}
\end{eqnarray}
where
\begin{equation}
E^i= R^NU^{2/(N-2)} e^{4a\phi/(N-1)}F^{0i}\,.
\label{eq18}
\end{equation}

Now we take the following ansatz for the dilatonic field:
\begin{equation}
e^{-4a\phi/(N-1)}=C(R(t))^p U^{q/(N-2)}
\label{eq19}
\end{equation}
where $C$ is a constant.

Substituting these ansatze into equations (\ref{eq16}) and (\ref{eq17}),
which come from the $(i,j)$ and $(0,i)$ components of the Einstein
equations, and the equation of motion for the electric field, we find
that $q$ must be $2a^2$ and it is required that
\begin{equation}
\frac{\partial^2}{\partial t\partial x^i}
((R(t))^{N-2+(p/2)}U^{(N-2+a^2)/(N-2)})=0\,.
\label{eq20}
\end{equation}

The other equations of motion require that
\begin{equation}
\nabla^2F=0\quad\mbox{(mod delta functions)}\,.
\label{eq21}
\end{equation}

Therefore we can choose
\begin{equation}
A_\mu dx^\mu=\sqrt{\frac{N-1}{2(N-2+a^2)}}
\frac{1}{\sqrt{C}(R(t))^{p/2}}(1-F^{-1}({\bf x},t))\, dt
\label{eq22}
\end{equation}
and
\begin{eqnarray}
F({\bf x},t)&\equiv& U^{(N-2+a^2)/(N-1)}({\bf x},t)\nonumber \\
&=&1+
\sum_{m=1}^n\frac{\mu_m}{\sqrt{C}(R(t))^{N-2+(p/2)}(N-2)|{\bf x}-{\bf
x}_m|^{N-2}}
\label{eq23}
\end{eqnarray}
for a solution describing $n$ (spinless) point sources located at
${\bf x}_m$ in cosmological background. Considering an isolated source
and the limit of small $\Lambda$ (i.e., very slow expansion), we find
that the mass, electnc charge and dilatonic charge of each point source
correspond to those of the source in the static background \cite{1}
\begin{eqnarray}
M_m&=&\frac{(N-1)A_{N-1}}{8\pi(N-2+a^2)}\mu_m
\label{eq24}\\
Q_m&=&\sqrt{\frac{N-1}{2(N-2+a^2)} }\mu_m
\label{eq25}\\
\sigma_m&=&\sqrt{\frac{N-1}{2}}\frac{a}{N-2+a^2}\mu_m
\label{eq26}
\end{eqnarray}
where
\begin{equation}
A_{N-1}=\frac{2\pi^{N/2}}{\Gamma(N/2)}
\label{eq27}
\end{equation}
when $R(t)=1$. (The explicit derivation of the relations
(\ref{eq24})-(\ref{eq26}) for the charged dilatonic black holes in the
static case can be found in
\cite{1}.) These relations satisfy the extremity condition for each
source
\cite{1}. Thus the sources are regarded as `extreme' black holes.

After these considerations, there remain three simultaneous, ordinary
differential equations with respect to the time variable. Apart from the
time-independent solution for $\Lambda=0$, which is obtained in
\cite{1}, one find only three cases below, in which consistent
solutions can be obtained as long as we adopt the above ansatz. The
three cases are:
\begin{eqnarray}
(I)\quad p&=&0\,; \nonumber \\
(II)\quad p&=&2a^2\,; \nonumber \\
(III)\quad p&=&-2(N-2)\,. \nonumber
\end{eqnarray}

In case (I), it is found that the solution which obeys the ansatz can
be obtained only if $a=b=0$. In this case, the solution reduces to:
\begin{eqnarray}
& &A_\mu dx^\mu=\sqrt{\frac{N-1}{2(N-2)}}(1-F^{-1}({\bf x},t))\,dt
\label{eq28}\\
& &\phi=constant
\label{eq29}\\
& &F({\bf x},t)=U({\bf x},t)=1+\sum_{m=1}^n
\frac{\mu_m}{(R(t))^{N-2}(N-2)|{\bf x}-{\bf x}_m|^{N-2}}
\label{eq30}\\
& &R(t)=e^{H(t-t_0)}
\label{eq31}
\end{eqnarray}
where $t_0$ is a constant and the constant $H$ satisfies
\begin{equation}
H^2=\frac{\Lambda}{N(N-1)}\,.
\label{eq32}
\end{equation}

This solution is no other than a straightforward extension of the
solution obtained in \cite{3} to $(N+1)$ dimensions. There is no
restriction on the value for the cosmological constant.

In case (II), it is found that the solution which obeys the ansatz can
be obtained only if $a=b$. In this case, the solution is reduced to
\begin{eqnarray}
& &A_\mu
dx^\mu=\sqrt{\frac{N-1}{2(N-2+a^2)}}\frac{1}{\sqrt{C}(R(t))^{a^2}}
(1-F^{-1}({\bf x},t))\,dt
\label{eq33}\\
& &e^{-4a\phi/(N-1)}=C(R(t))^{2a^2}U^{2a^2/(N-2)}
\label{eq34}\\
& &F({\bf x},t)=U^{(N-2+a^2)/(N-2)}({\bf x},t)\nonumber \\
& &\qquad\quad=1+\sum_{m=1}^n
\frac{\mu_m}{\sqrt{C}(R(t))^{N-2+a^2}(N-2)|{\bf x}-{\bf x}_m|^{N-2}}
\label{eq35}\\
& &R(t)=(t/t_0)^{1/a^2}
\label{eq36}
\end{eqnarray}
with a relation
\begin{equation}
\frac{1}{a^2}\left(\frac{N}{a^2}-1\right)=\frac{C^{-1}t_0^2\Lambda}{N-1}\,.
\label{eq37}
\end{equation}

One can verify that the scale factor obeys the expansion law of the
vacant space, for $a=b$ (i.e., equations (\ref{eq36}), (\ref{eq34}),
(\ref{eq37}) are reduced to (\ref{eq9}), (\ref{eq10}), (\ref{eq11}) when
all $\mu_m=0$ for $a=b$). There is a critical value at
$a^2=N$; $\Lambda$ takes a positive value of $a^2<N$, while $\Lambda$
must be negative if $a^2>N$.

The action for the effective field theory of string theory corresponds
to the action (\ref{eq1}) with $a=b=1$. This case is included in the
case (II). If we can rescale the metric as $\bar{g}_{^mu\nu}=
e^{4a\phi/(N-1)}g_{\mu\nu}$ for string theory, the action becomes
\begin{equation}
S=\int d^{N+1}x\, \frac{\sqrt{-\bar{g}}}{16\pi}e^{-2\phi}\left[
\bar{R}+4(\bar{\nabla}\phi)^2-\bar{F}^2-\Lambda\right]\,.
\label{eq38}
\end{equation}

In terms of the string metric $\bar{g}_{\mu\nu}$, the cosmological
metric we have obtained is
\begin{equation}
d\bar{s}^2=C^{-1}[-F^{-2}({\bf x},\tau) d\tau^2+d{\bf x}^2]            
\label{eq39}
\end{equation}
and the dilaton field is given by
\begin{equation}
e^{-4\phi/(N-1)}=CF^{2/(N-1)}e^{\pm 2\tau/t_0}
\label{eq40}
\end{equation}
where
\begin{equation}
F({\bf x},\tau)=1+
\frac{\mu_m}{\sqrt{C}(N-2)e^{\pm(N-1)(\tau/t_0)}|{\bf x}-{\bf x}
_m|^{N-2}}
\label{eq41}
\end{equation}
and
\begin{equation}
\tau\equiv\pm t_0\ln(t/t_0)\,.
\label{eq42}
\end{equation}

If all $\mu$ are zero, the solution (for $a=b=1$) reduces to the
linear-dilaton background \cite{6} in terms of the string metric.

In case (III), it is found that the solution which obeys the ansaz
can be obtained only if the relation $ab=-(N-2)$ holds. In this
case, the solution reduces to: 
\begin{eqnarray}
& &A_\mu
dx^\mu=\sqrt{\frac{N-1}{2(N-2+a^2)}}\frac{(R(t))^{(N-2)/2}}{\sqrt{C}}
(1-F^{-1}({\bf x}))\,dt
\label{eq43}\\
& &e^{-4a\phi/(N-1)}=C(R(t))^{-2(N-2)}U^{2a^2/(N-2)}
\label{eq44}\\
& &F({\bf x})=U^{(N-2+a^2)/(N-2)}({\bf x})\nonumber \\
& &\qquad\quad=1+\sum_{m=1}^n
\frac{\mu_m}{\sqrt{C}(N-2)|{\bf x}-{\bf x}_m|^{N-2}}
\label{eq45}\\
& &R(t)=(t/t_0)^{a^2/(N-2)^2}
\label{eq46}
\end{eqnarray}
with a relation
\begin{equation}
\frac{a^2}{(N-2)^2}\left(\frac{Na^2}{(N-2)^2}-1\right)=
\frac{C^{(N-2)/a^2}t_0^2\Lambda}{N-1}\,.
\label{eq47}
\end{equation}

Again in this case, the scale factor is found to obey the expansion
law of the vacant space, for $a=-(N-2)/b$ (i.e., equations (\ref{eq46}),
(\ref{eq44}), (\ref{eq47}) are reduced to (\ref{eq9}), (\ref{eq10}),
(\ref{eq11}) when all $\mu_m=0$ for
$a=-(N-2)/b)$. There is a critical value at $a^2=(N-2)^2/N$; The value
of $\Lambda$ must be negative if $a^2<(N-2)^2/N$, while $\Lambda$ is
positive if $a^2>(N-2)^2/N$.

In the Einstein-Maxwell-dilaton system which is governed by the action
(\ref{eq1}), these three cases give all of the explicit solutions
derived from the ansatze (\ref{eq12}) and (\ref{eq19}) in the presence
of point charges in cosmological background.

Now, in order to investigate the nature of the spacetime geometry,
we examine propagation of light in the metric describing one extreme
black hole in a cosmological background.

 We consider the light emitted in the radial direction out- and
inward from a source located at $r=r_0$ at the time $t=t_0$, when
the scale factor is normalized to unity. Because the propagation of the
light follows the null geodesic, it is found that
\begin{equation}
\frac{dr}{dt}=\pm\frac{1}{R(t)U^{(N-1)/(N-2)}(r,t)}
\label{eq48}
\end{equation}
where
\begin{equation}
U(r,t)=\left(1+
\frac{\mu}{\sqrt{C}(R(t))^{N-2+(p/2)}(N-2)r^{N-2}}
\right)^{(N-2)/(N-2+a^2)}
\label{eq49}
\end{equation}
and $R(t)$ is normalized as $R(t_0)=1$. In the right-hand side of
equation (\ref{eq48}), a positive sign means outward propagation, while
a negative one means inward.

For the case without the cosmological constant, the trajectory of the
light is described by the equations (\ref{eq48}), (\ref{eq49}) setting
$R\equiv 1$. In this static case, the outward light from the point
which is located at $r_0>0$ reaches an arbitrary distance in a finite
time interval. If $r_0$ approaches zero, the time becomes infinitely
long. Then we can guess that the honzon is located at $r=0$ in this
coordinate. No other trapped surface is found as long as the
background metric is static in this case.

Let us look into the light propagation for each solution with the
cosmological constant, which has been obtained previously by the author
of \cite{3}.

Case (I) ($a=b=0$). The solution for the light propagation is given
by
\begin{equation}
t-t_0=\frac{1}{N-2}\int_{r_0^{N-2}}^{(Rr)^{N-2}}\frac{dY}{Y}
\frac{[Y+\mu/(\sqrt{C}(N-2))]^{(N-1)/(N-2)}}{H[Y+
\mu/(\sqrt{C}(N-2))]^{(N-1)/(N-2)}\pm
Y}\,.
\label{eq50}
\end{equation}

For inward propagation, the light from any distance never reaches a
certain distance $r_i$ in a finite time interval. It may be said that an
inner trapped surface is located at $r=r_i$ \cite{3}. Therefore, this
metric actually describes a white hole \cite{3}. For $N=3$, $r_i$ is
given by
\begin{equation}
r_i=\frac{1}{2HR}\left(\left(1-\frac{2\mu H}{\sqrt{C}}\right)-
\sqrt{1-\frac{4\mu H}{\sqrt{C}}}\right)\,.
\label{eq51}
\end{equation}

Note that the physical length $Rr_i$ is constant.

Also, for inward propagation, the light emitted from a source at
$r_0>r_H$ never reaches a certain interior region ($r<r_H$). It can be
said that the de Sitter horizon is located at $r=r_H$
\cite{3}. For $N=3$, $r_H$ is given by
\begin{equation}
r_H=\frac{1}{2H}\left(\left(1-\frac{2\mu H}{\sqrt{C}}\right)+
\sqrt{1-\frac{4\mu H}{\sqrt{C}}}\right)\,.
\label{eq52}
\end{equation}

This horizon length reduces to the well known value $H^{-1}$, when
$\mu=0$.

Case (II) ($a=b$). We have not yet found explicit solutions for
arbitrary values of $a$. The solutions are found, however, for $a=1$.

The solution for the light propagation for the $a=b=1$ case is
given by
\begin{equation}
t^{N-1}= e^{\pm(N-1)(r-r_0)}t_0^{N-1}\pm\frac{(N-1)\mu}{(N-2)\sqrt{C}}
e^{\pm(N-1)r}\int_{r_0}^{r}\frac{dr}{e^{\pm(N-1)r}r^{N-2}}\,.
\label{eq53}
\end{equation}

For outward propagation, the light from $r_0= 0$ never reaches a finite
distance in a finite time interval. Thus the black hole horizon is
located at $r_+=0$ in this coordinate. The location of the black hole
horizon turns out to be unchanged by the effect of the cosmological
constant. There is no cosmological (event) horizon for $a=1$ in this
case.

Case (III) ($ab=-(N-2)$). The solution for the light propagation is
given by
\begin{equation}
\int_{t_0}^t\frac{d\tilde{t}}{R(\tilde{t})}=\pm\int_{r_0}^{r}
dr\left(1+\frac{\mu}{\sqrt{C}(N-2)r^{N-2}}\right)^{(N-1)/(N-2+a^2)}
\label{eq54}
\end{equation}

For outward propagation, the light from $r_0=0$ never reaches a finite
distance in a finite time interval. Thus the black hole horizon is
located at $r_+=0$ in this coordinate.
This is unchanged from the case with vanishing cosmological constant.

For inward propagation, it happens that the light emitted from a
source at $r_0$ cannot reach a point located at $r<r_{min}$. This occurs
when $a^2>(N-2)^2$, or equivalently, $b^2<1$;
\begin{equation}
\frac{1}{t_0^{(1/b^2)-1}}=\left(\frac{1}{b^2}-1\right)
\int_{r_{min}}^{r_0}\left(1+\frac{\mu}{\sqrt{C}(N-2)r^{N-2}}
\right)^{(N-1)/(N-2+a^2)}~(b^2<1)\,. 
\label{eq55}
\end{equation}

Further properties of the spacetime will be studied in a separate
publication.

In conclusion, we have found cosmological multi-black (white) hole
solutions in Einstein-Maxwell-dilaton theory and analysed the
propagation of light in the background of cosmological one-black (white)
hole solution.

The critical values for the coupling $a$ in cases (II) and (III), which
determine the sign of the cosmological constant, coincide with the
critical values found in the analysis of the low-energy scattering of
black holes when the cosmological constant vanishes \cite{1}; when $a^2$
is larger than
$(N-2)^2/N$, coalescence of black holes does not occur; when $a^2=N$,
there is no interaction  between two black holes. We have to study
the coincidence more closely.

The relation to supersymmetry is still unclear. The $N=4$
supergravity including vector fields and a cosmological constant is
known to correspond to the action (\ref{eq1}) with
$a=b=1$ \cite{7}. But the sign of the cosmologlcal constant is
negative as for ordinary gauge theories. Unless we should consider an
imaginary gauge coupling or a ghost-like vector field, we cannot apply
our background solution to such supergravity theory.

A possiblity remains in some supergravity theory of which the
background is partially compactified with and without uniform
electromagnetic field. One might look for multi-black hole solution
in the `electrovac' \cite{7} background.

Finally, one may ask the questions: `Can we learn something about the
early stages of the big-bang by studying these solutions?' and `Is a
massive point source a string excitation?' We confess that it is
difficult to answer these questions in this paper. For the concrete
application to cosmology, we must consider other matter fields (possibly
at finite temperature). Moreover we may take the other features of
string tbeory into consideration. The connection to string theory in
non-critical dimensions is very important, for our solution for $a=
b=1$ describes all arbitrary number of point charges in the linear
dilaton background \cite{6}. We must study the `stringy' interpretation
of the background and sources in detail. Inclusion of other
antisymmetric tensor fields may also be worth studying.%
\footnote{%
Multi-centered solutions in the effective theory or string theory are
further investigated separately in \cite{8}.
} Further studies will be presented elsewhere in the future.

\section*{Acknowledgment}
The authors would like to thank the referees for some enlightening
comments.

%%%%%%%%%%%%%%%%%%%%%%%%%%%%%%%%%%%%%%%%%%%%%% 

%%%%%%%%%%%%%%%%%%%%%%%%%%%%%%%%%%%%%%%%%%%%%
\end{document}